\documentclass{iaus}
\usepackage{lscape}
\usepackage{rotating}
\usepackage{times}
\def\Msun{\hbox{M$_\odot$}}

\def\cm3{\hbox{cm$^{-3}$}}

\newcommand\hst{{\it HST}}
\newcommand\ie{i.\,e.~}

\newcommand\dndt{${\textup{d}N/}\textup{d}\tau$}
\newcommand\ubvi{\textit{UBVI}}

\usepackage{graphicx}
\usepackage{amssymb}
\title[IAUS 266~~Star cluster disruption mechanisms]{Constraining star cluster disruption mechanisms}
\author[I.~S.~Konstantopoulos et al.]{I.~S.~Konstantopoulos$^{1,2,*}$, 
N.~Bastian$^3$, 
M.~Gieles$^2$, 
\and H.~J.~G.~L.~M.~Lamers$^4$}

\affiliation{$^1$ Department of Physics \& 
Astronomy, University College London\\ Gower Street, London, WC1E 6BT, UK\\
$^2$ European Southern Observatory, Casilla 19001, Santiago~19, Chile\\
$^3$ Institute of Astronomy, University of Cambridge, Madingley Road, Cambridge, CB3 0HA, UK\\
$^4$ Astronomical Institute, Utrecht University, Princetonplein 5, 3584 CC Utrecht, The Netherlands\\
$^*$ Currently at Penn State University; email: {iraklis@astro.psu.edu}
}

\pubyear{2010}
\volume{266}
\pagerange{1--2}
\jname{Star Clusters: Basic Galactic Building Blocks Throughout Time And Space}
\editors{R.~de~Grijs and J.~Lepine Editors}

\begin{document}

\maketitle

\begin{abstract}
Star clusters are found in all sorts of environments, and their formation and evolution is inextricably linked to the star formation process. Their eventual destruction can result from a number of factors at different times, but the process can be investigated as a whole through the study of the cluster age distribution. 
Observations of populous cluster samples reveal a distribution following a power law of index approximately $-1$. In this work we use M33 as a test case to examine the age distribution of an archetypal cluster population and show that it is in fact the evolving shape of the mass detection limit that defines this trend. That is to say, any \textit{magnitude-limited} sample will appear to follow a $dN/d\tau=\tau^{-1}$, while cutting the sample according to mass gives rise to a composite structure, perhaps implying a dependence of the cluster disruption process on mass. In the context of this framework, we examine different models of cluster disruption from both theoretical and observational standpoints. 
\keywords{galaxies: star clusters, galaxies: individual M33}
\end{abstract}

\firstsection
\section{Introduction}\label{intro}
Star clusters are commonly used to trace the stellar content 
and star formation history of their host systems. The 
main limitations to this approach are the finite lifetime of 
a cluster (disruption) and evolutionary fading. 
From a theoretical point of view, the lifetime of a cluster 
should depend upon its initial mass and the properties of 
the environment in which it evolves (\cite[Spitzer 1958]{spitzer58}\footnote{This work showed that the lifetime due to encounters with interstellar clouds depends on the cluster density. Observations of young clusters have shown the radius to scatter tightly about 3.5~pc and can thus be considered a constant for any practical formulation. This immediately establishes the cluster mass as the main deciding factor of the lifetime of a cluster.}; \cite[Baumgardt \& Makino 2003]{bm03}). Observations have, 
however, produced \textit{two} empirical disruption laws: 
\begin{itemize}
	\item Mass dependent disruption (MDD, \cite[Boutloukos \& Lamers 2003]{bl03})
	\item Mass independent disruption (MID, \cite[Fall, Chandar \& Whitmore 2005]{fcw05}) 
\end{itemize}

In the MID model, `survivors' are selected on a purely 
random basis and a constant fraction is destroyed every 
age dex. Intriguing as it may be, this model clashes with 
several principles of cluster dynamics that would need to 
be revised considerably to accommodate it. In this contribution 
we test cluster dissolution and attempt to disentangle 
it from incompleteness and the statistical biases that have 
in the past limited such studies. We then compare 
theory to observations of the M33 cluster system.

\section{A theoretical preamble}\label{sec:theory}
\begin{figure}
\begin{center}
\includegraphics[width=0.45\textwidth]{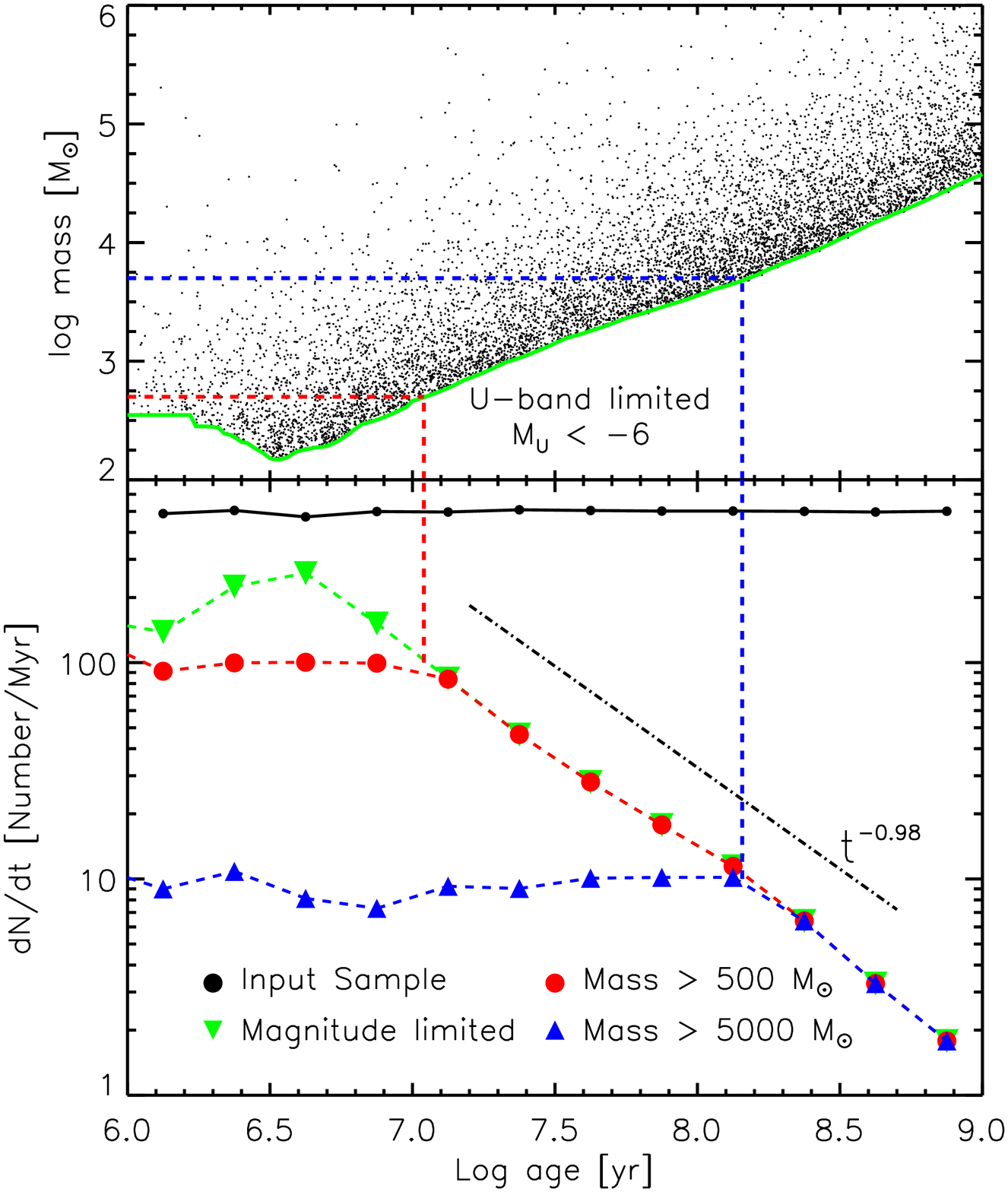}~ \includegraphics[width=0.45\textwidth]{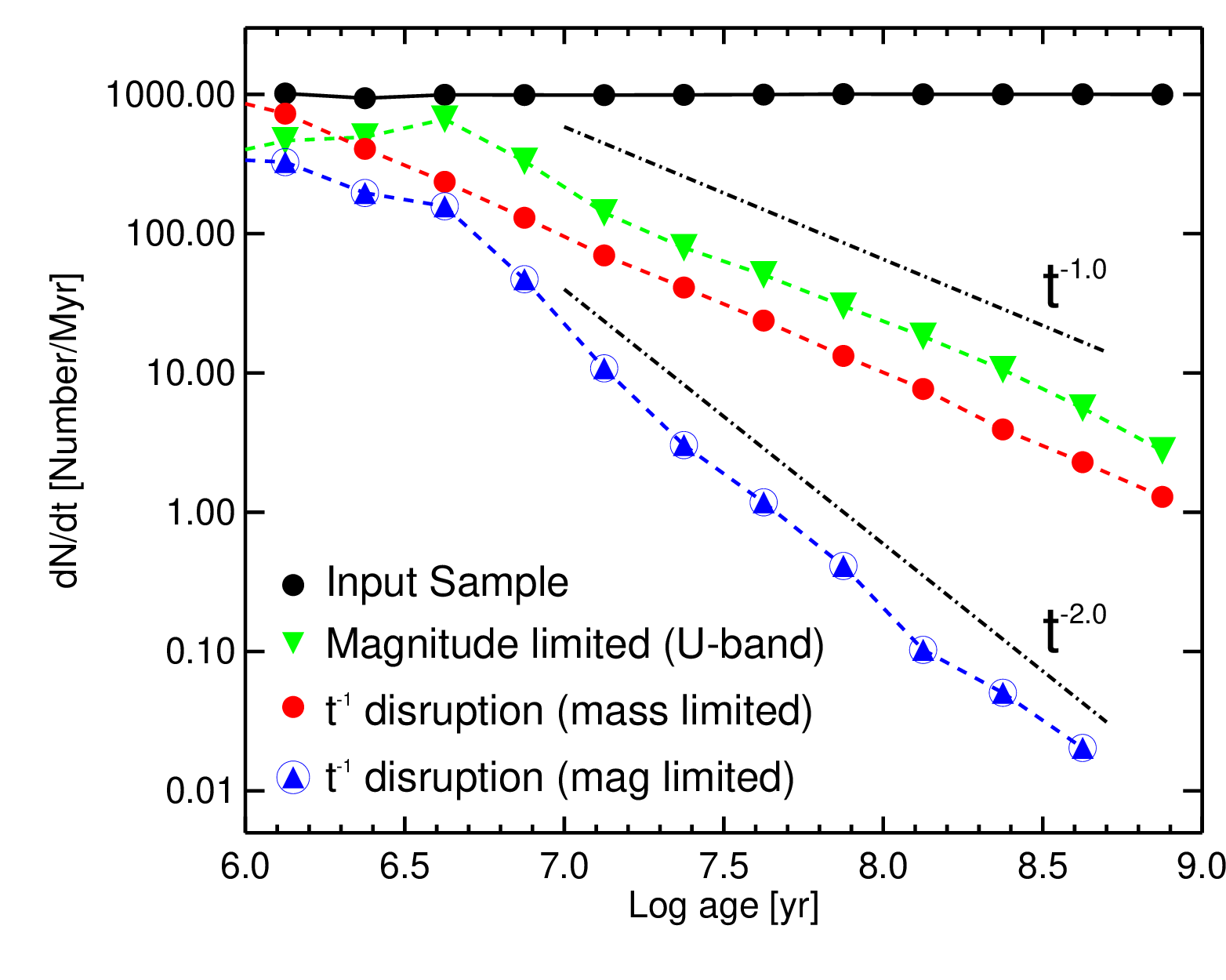}
\caption{{\bf Left:}~Analysis of an artificial cluster population at the distance of M33. The population contains $5\times10^5$ clusters, sampled from a power-law mass function with index $-2$, and assuming a constant cluster formation rate. The {\bf top panel} shows the mass distribution for the observable part of the population -- the detection limit is denoted by the solid line. This takes into account evolutionary fading, and the sample is assumed to be limited by the $U$-band, as is most commonly the case in extragalactic studies. The {\bf bottom panel} shows the number of clusters per logarithmic age bin (\dndt): the dark solid line corresponds to all clusters `formed' (\ie\ the entire simulated set), while the dashed green line tracks `observed' clusters, \ie\ those above the detection limit. The theoretical expectation is for a magnitude limited sample to follow a simple power law distribution, whereas the introduction of a mass limit segments the distribution into a flat and a power law part (the transition is dictated by the mass cut and the point where incompleteness sets in). {\bf Right:}~A similar population, according to the predictions of MID. This presents the age distribution of a cluster population with a constant cluster formation rate and an imposed a $\tau^{-1}$ disruption law.
}
\label{fig:theory}
\end{center}
\end{figure}

Before providing a treatment of an observed data-set, 
let us try to understand the `tools of the 
trade', the statistical distributions used to study 
and characterise the cluster disruption process. 
To achieve that, we have created an artificial 
population with a constant cluster formation 
history (CFH) and masses drawn from a power-law 
distribution with index $-2$ -- although we note that the cluster initial mass function was recently found by \cite[Larsen~(2009)]{larsen09} and \cite[Gieles~(2009)]{gieles09} to be better described by a Schechter function. This population is presented in the left panel of Fig.~\ref{fig:theory}. 

\subsection{Age-mass diagram}
This plot simply shows the number of clusters 
with increasing age. Log-bins accentuate the increase 
of the number of high mass clusters with increasing age. 
This arises because the sample grows with time 
-- a `size-of-sample' effect. 
As the sample increases, so does the 
likelihood of producing a massive cluster. 

\subsection{Detection limit}
Unlike the top envelope of the age-mass diagram, 
the line along the bottom of the distribution is 
not caused by statistics, but the detection limit 
of our modelled observations. As clusters fade with 
time, this constant limiting flux translates into 
an increasing limiting mass. 

In most of the cases observed cluster samples will be luminosity limited. 
As the population ages, 
clusters of higher mass fade below the detection 
limit. We have denoted this by a green line, 
representing the mass of a cluster with $M_U=-6$ 
(adapted to the study of M33 that will follow) at 
a given age. $U$~band is normally the limiting 
filter in observational studies and we emulate the 
detection limit in this plot. 

\subsection{Age distribution -- the \dndt\ plot}
The lower panel shows the main diagnostic 
used in the study of cluster disruption. 
The vertical axis gives the number of clusters our 
imaginary galaxy formed in each log-age bin, 
normalised to unit time (in this case one Myr). 
Here we present the input sample as a solid 
line: a constant CFH gives rise to a flat 
line. This will not, however, be observed, 
due to the detection limit.

\subsection{A magnitude-limited sample}
The green line shows all clusters that lie above the 
detection limit in each log-age bin and displays 
the characteristic power-law shape found in all 
observational studies. The interpretation of this plot 
is far from trivial: this $\tau^{-1}$ shape means that 90\% of 
the population is lost with each age dex. MID 
interprets this as 90\% of clusters dissolving each 
age dex (\eg\ in M33, \cite[Sarajedini \& Mancone 2007]{sm07}). 
In this example, however, this is due to 
detection incompleteness. Thus, a model has to 
treat \textit{fading} and \textit{dissolution} as 
two separate and concurrent causes of cluster disappearance. 

\subsection{Taking mass cuts}
Having established the above, we can now perform a 
simple test for the mass dependence of cluster disruption: 
cutting the sample according to mass. 

In the \dndt\ plot we present two mass cuts as a red and blue 
line respectively. Both cuts split the power-law distribution to a 
composite shape: a flat part, i.e. a constant cluster 
formation rate, and a power-law decline, due to clusters 
fading below the detection limit.

\section{Observational study}\label{sec:obs}
After understanding the caveats inherent in the 
study of the \dndt\ plot, we can proceed to plot 
the distribution of an observed cluster sample. 

\begin{figure}
\begin{center}
\includegraphics[width=0.45\textwidth]{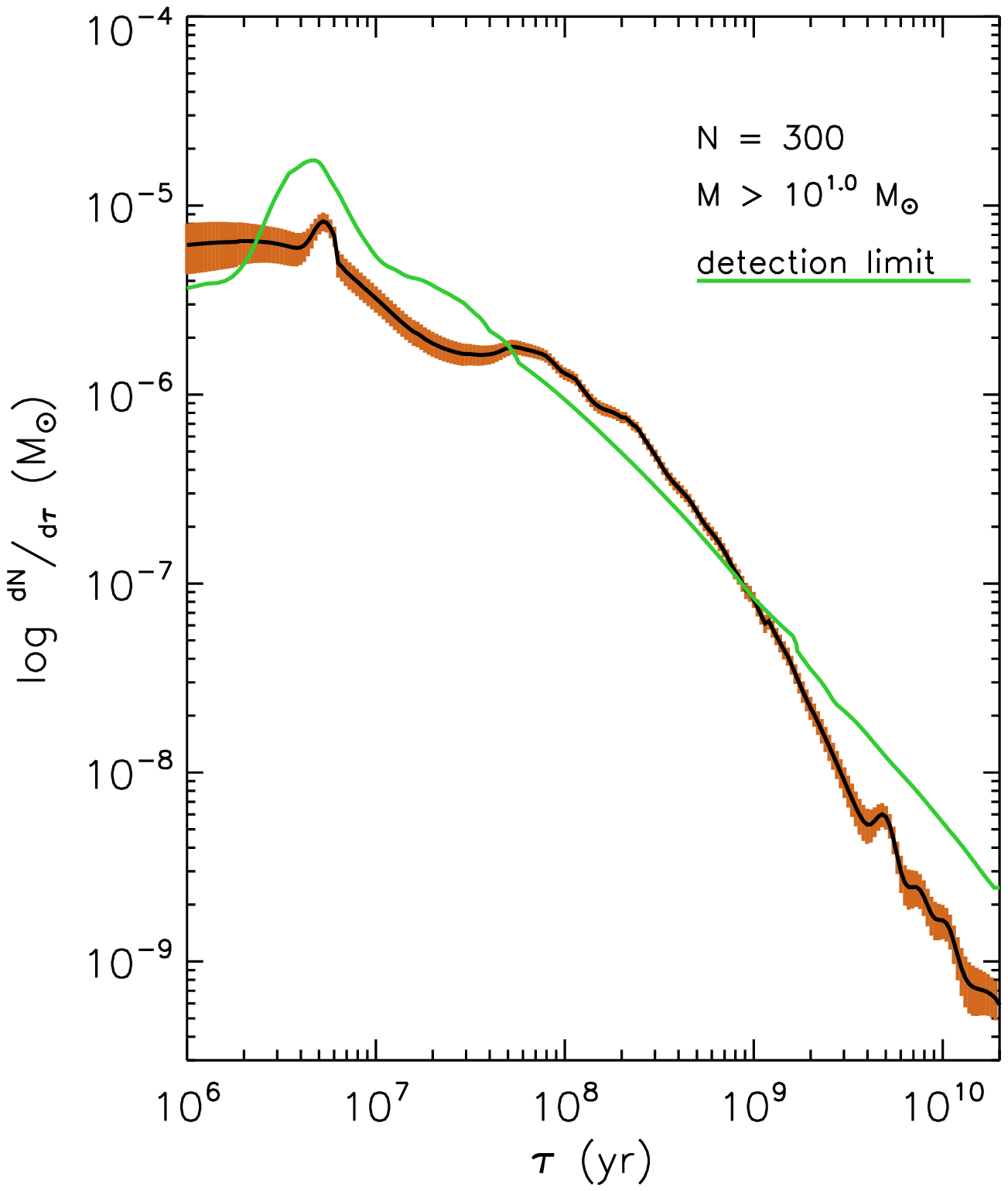}~\includegraphics[width=0.45\textwidth]{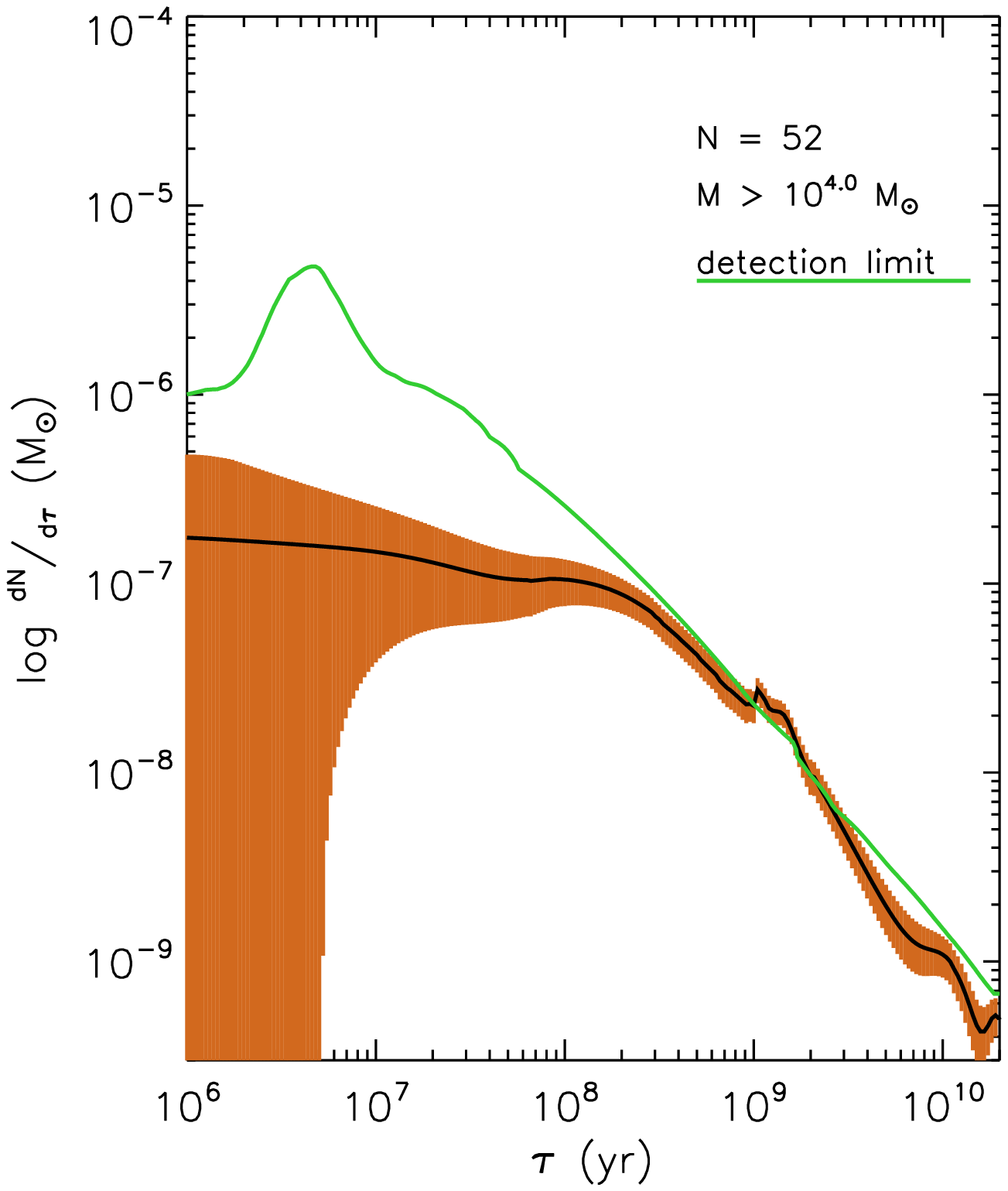}
\caption{{\bf Left panel:}~A `smoothed' \dndt\ plot: here we represent each measurement as a Gaussian, where the peak occurs at the best fit value (from 3DEF) and the wings are shaped after the uncertainty in the age fit (defined as the extrema calculated by 3DEF). This minimises the effect of uncertain measurements on the shape of the distribution, as an erratic measurement will be represented by a flat Gaussian. The solid green line is a prediction of the \dndt\ of a luminosity limited sample. It is the detection limit of Fig.~\ref{fig:theory} (left), inverted to reflect the maximum expected number of detections, given the detection limit (due to the complex scaling of the \dndt\ plot we have normalised the two lines at 1~Gyr). {\bf Right panel:}~the same plot with an imposed mass cut at $10^4$~\Msun. The obvious flattening with respect to the magnitude limited sample on the left strongly supports a mass dependence in the cluster disruption process (consistent with BL03).}
\label{fig:obs}
\end{center}
\end{figure}

We obtained the age and mass of a sample of 
$\sim350$ \textit{bona fide} clusters in M33 (spanning the 
entire surface of the galaxy, an unbiased, 
\hst-selected sample; \cite[Sarajedini \& Mancone 2007, San Roman et~al.~2009]{sm07,sr09}) using \ubvi\ imaging 
from the Local Group Survey (\cite[Massey~et~al.~2006]{m06}). The SFR is 
known to have been constant for $\sim1$~Gyr in 
M33, making it ideal for our study. Fig.~\ref{fig:obs} 
presents the age distribution of the magnitude-limited 
sample (left) and a mass cut (right). It 
exhibits the same shape as the \dndt\ plot of 
Fig.~\ref{fig:theory} (left), with a flat part leading to a power-law.

We also provide a direct test of MID in Fig.~\ref{fig:theory} (right panel), 
where we create a population that destroys 90\% 
of its clusters each age dex \textit{and} is subject to 
evolutionary fading. 

\subsection{Observed \dndt\ plot}
In order to dampen the effect of local overdensities in the 
\dndt\ plot (caused by uncertainties in the SSP models), 
in Fig. 2 we represent 
each age by a Gaussian, where the 
wings are defined by the uncertainty. 
The result is consistent with a  
$\tau^{-1}$ power-law for a magnitude-limited 
sample. 

Crucially, as predicted by the models 
of Fig.~\ref{fig:theory} (left), taking a mass cut produces a 
composite structure, with a flat initial 
part and a power-law decline at older 
ages. 

This demonstrates that an observed 
sample needs to be interpreted as the 
result of fading and dissolution at different timescales. 

\subsection{Mass independent dissolution plus fading}
We have argued so far that the observed \dndt\ 
plot will show a combination of disruption and fading. 
The right hand side panel of Fig.~\ref{fig:theory} shows the combined effect of 
90\% dissolution (the MID hypothesis) \textit{and} 90\% 
disappearance due to fading. This results in a very steep 
power law decline (blue line) that is \textit{inconsistent} 
with the observations of Fig.~\ref{fig:obs}. 

This result stands firmly against the MID 
cluster disruption model and implies the existence of a mass 
dependence.

\section{Summary}\label{sec:summary}
Sample incompleteness causes a $\tau^{-1}$ decline in the observed age 
distribution of cluster populations. Size-of-sample effects and 
thorough statistical methods can help to interpret the disruption 
process at play. The fundamental difference between magnitude- 
and mass-limited samples can then be used to discover the underlying physics. Our results from the nearby population of M33 are 
inconsistent with mass-independent disruption of a fixed fraction of clusters per age dex.

\section*{Acknowledgements}
ISK gratefully acknowledges the support of an ESO studentship for the undertaking of this work and would like to thank the Vitacura staff and students for their hospitality. Thanks are also due to the organising committee of the General Assembly and Symposium 266. \textit{Obrigado Rio!}


\begin{thebibliography}{}
\bibitem[Baumgardt \& Makino(2003)]{bm03} Baumgardt, H., \& Makino, J.\ 2003, MNRAS, 340, 227 
\bibitem[Boutloukos \& Lamers(2003)]{bl03} Boutloukos, S.~G., \& Lamers, H.~J.~G.~L.~M.\ 2003, \textit{MNRAS}, 338, 717 
\bibitem[Fall et al.(2005)]{fcw05} Fall, S.~M., Chandar, R., \& Whitmore, B.~C.\ 2005, \textit{ApJL}, 631, L133 
\bibitem[Larsen(2009)]{larsen09} Larsen, S.~S.\ 2009, \textit{A\&A}, 503, 467 
\bibitem[Gieles(2009)]{gieles09} Gieles, M.\ 2009, \textit{MNRAS}, 394, 2113 
\bibitem[Sarajedini \& Mancone(2007)]{sm07} Sarajedini, A., \& Mancone, C.~L.\ 2007, \textit{AJ}, 134, 447 
\bibitem[San Roman et al.(2009)]{sr09} San Roman, I., Sarajedini, A., Garnett, D.~R., \& Holtzman, J.~A.\ 2009, \textit{ApJ}, 699, 839 
\bibitem[Massey et al.(2006)]{m06} Massey, P., Olsen, K.~A.~G., Hodge, P.~W., Strong, S.~B., Jacoby, G.~H., Schlingman, W., \& Smith, R.~C.\ 2006, \textit{AJ}, 131, 2478 
\bibitem[Spitzer(1958)]{spitzer58} Spitzer, L.~J.\ 1958, ApJ, 
127, 17 

\end{thebibliography}
\end{document}